\documentclass[]{mn2e}
\usepackage{graphicx}	% Including figure files
\usepackage{amsmath}	% Advanced maths commands
\usepackage{amssymb}	% Extra maths symbols 
\usepackage{bm}
\def\apj{ApJ} \def\apjs{ApJS}\def\mnras{MNRAS}\def\pasj{PASJ}\def\nat{Nat.}\def\aap{AA}

\def\kms{km s$^{-1}$} 
\def\co{$^{12}{\rm CO}(J=1-0)$}
\def\vlsr{V_{\rm lsr}}    \def\Msun{M_\odot} 
\def\deg{^\circ} \def\Tb{T_{\rm B}}  
\def\Ico{I_{\rm CO}}

 \def\/{\over}
\def\be{\begin{equation}} \def\ee{\end{equation}}

\def\({\left(} \def\){\right)} \def\[{\left[} \def\]{\right]}
\def\be{\begin{equation}} \def\ee{\end{equation}}

 \def\({\left(} \def\){\right)}
\def\S{\mathcal{S}} 
\def\ergcc{erg cm$^{-3}$}
\def\ergccs{\ergcc s$^{-1}$}
\def\muG{$\mu$G }

\def\OmegaGSB{ $\Omega$/{\rotatebox[origin=c]{180}{$\Omega$}} } 

\title[Galactic-Center Super Bubbles]
{Interaction of the Galactic-Centre Super Bubbles with the Gaseous Disc}  
\author[Y. Sofue and J. Kataoka]{Yoshiaki {\sc Sofue}$^{1}$\thanks{E-mail: sofue@ioa.s.u-tokyo.ac.jp} and Jun {\sc Kataoka}$^2$ \\
1. Institute of Astronomy, The University of Tokyo, Mitaka, Tokyo 181-0015, Japan \\
2. Faculty of Science and Engineering, Waseda University, Shinjyuku, Tokyo, 169-8555, Japan
} 
   
\begin{document}   
\maketitle

\begin{abstract}  
The interaction of Galactic-Centre (GC) super bubbles (GSB) with the gaseous disc and halo of the Milky Way is investigated using radio continuum, X-ray, HI and CO line surveys.
The radio North Polar Spur (NPS) constitutes the brightest eastern ridge of GSB, brightening towards the galactic plane and reaching  $ l = 22\deg, \ b = + 2\deg$ at the sharpest end, where it intersects the tangential direction of the 3-kpc expanding ring and crater.
Examination of the spur ridges reveals that the entire GSB, including the NPS and its counter spurs, constitutes a GC-symmetrical $\Omega /$\rotatebox[origin=c]{180}{$\Omega$} shape.
The thickness and gas density of the HI and CO discs are shown to increase sharply from the inside (lower longitude) to the outside of the 3-kpc crater.
Formation of crater is explained by the sweeping of the upper layer of disc gas by the shock wave from the GC by the explosion $ \sim 10 $ My ago with the emitted energy of several $10 ^ {55} $ ergs.
Based on the discussion, a unified view on the structure and formation mechanism of GSB is presented.  
\end{abstract}
 

%%%%%%%%%%%%%%%%%%%%%%%%%
\section{Introduction} 

The Galactic Centre (GC) exhibits a variety of explosive phenomena, associated with high-energy outflows and shock waves. 
The most energetic and gigantic shock wave so far reported is the GC super bubbles (GSB)\footnote{The same object has been given various names such as north/south polar spurs (NPS/SPS); radio loops; Galactic super bubbles; bipolar hyper shells (BHS); ROSAT/eROSITA bubbles, etc.. }
due to a nuclear explosion with released energy on the order of $\sim 10^{55-56}$ erg and extent as large as $\sim \pm 10$ kpc above and below the Galactic plane
(Sofue1977, 2000; Sofue et al 2016; Bland-Hauthorn et al. 2003; Crocker et al. 2015; Sakar et al. 2016; Kataoka et al. 2018). 
 In this paper, we revisit this prominent object in the Galaxy in view of the GC explosion model, being inspired by the recent excellent X-ray observations of the whole sky with the eROSITA \cite{predehl+2020}. 

High-altitude, large scale bubbles in the halo around the GC have been evidenced in the last decades by the all sky surveys in radio continuum emission \cite{haslam+1982,reich+2001,planck+2013}, X-rays \cite{snowden+1997,predehl+2020},  and $\gamma$-rays \cite{su+2010}, although their scale size and morphology differ in different energy bands. Moreover, giant magnetized outflows from the GC are also suggested by linear polarization measurement in radio band \cite{carretti+2013}.
 
The bipolar double horn structure has been interpreted as a shock wave from the GC, and is well simulated by the bipolar-hyper shell (BHS) model
{\cite{sofue1994,sofue2000,guo2012,sofue+2016,sarkar+2016,sarkar2019}}.
The connection of the GSB to the energetic phenomena in the GC has been firmly confirmed by the discovery of the Fermi bubbles
\cite{su+2010,crocker+2015,kataoka+2018}.
In fact, various outflow phenomena have been known in the GC region, such as the GC lobe (GCL) \cite{sofue+1984}
as recently revealed of its vertical connection to higher latitude halo \cite{heiw+2019}.
An X-ray chimney extending hundreds of parsecs above and below the GC has been discussed with tight correlation with the radio and infrared (IR) vertical structures \cite{nakashima+2013,ponti+2019,ponti+2021}.
Disturbed ionized gas appears to be flowing out at high speed from the GC into the halo below the galactic plane \cite{savage+2017}.
These observations seem to connect Sgr A to the Fermi bubbles, further corroborating the GC-origin scenario of the outer bubbles.
Interconnections and coherent features in radio, IR, X- and $\gamma$-ray maps are thus suggested in the central few degrees of the GC to $\sim 10$ kpc of the Galactic halo in the past $\sim 10-20$ My.
 
\begin{figure} 
\begin{center}  
\includegraphics[width=5.5cm]{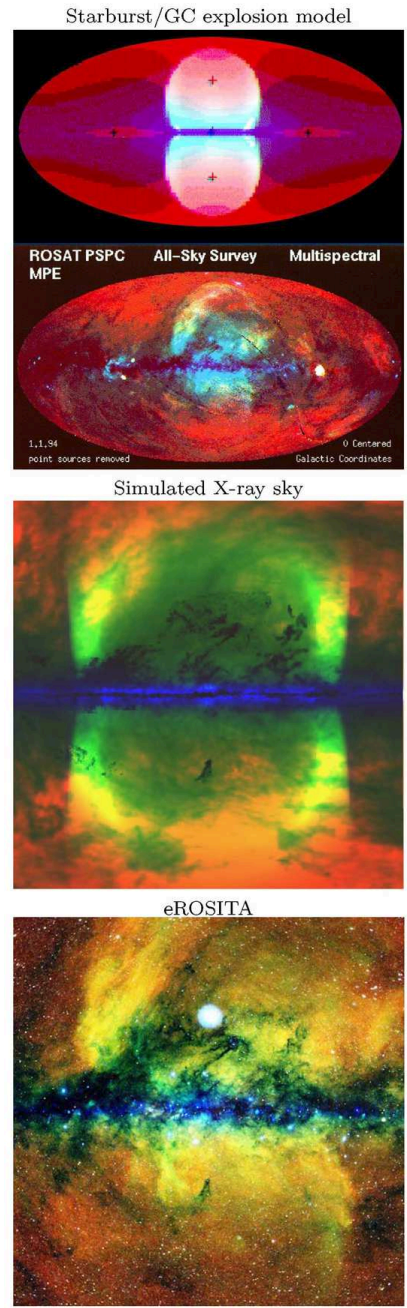}  
\end{center} 
\caption{
Summary of the bipolar hyper-shell model. 
[Top] Calculated X-ray sky (Sofue 2000) compared with the ROSAT all sky X-ray map (Ssnowden et al. 1997). RGB colors stand for 1/4, 3/4, and 1.5 keV, respectively. For intensities see the cited papers.
[Middle] Simulated X-ray map on the $100\deg\times 100\deg$  sky (RGB for 0.25, 0.75, and 2 keV in arbitrary intensity scaling) corrected for extinction (Sofue et al. 2016) using the HI survey data (Kalberla et al. 2005).
[Bottom] eROSITA RGB map of the central $100\deg\times 100\deg$ region (Predehl et al. 2020) as reproduced from url https://www.mpe.mpg.de/7461761/news20200619. Note the significant absorption near the galactic plane, where the radio data are crucial in order to see the interaction of the spurs with the disc. 
}
\label{fig1}  
\end{figure}

Fig. \ref{fig1} summarizes the starburst or GC explosion model \cite{sofue2000,sofue+2016}, which considered the shock wave propagation in the disc and halo, as compared with the X-ray  all-sky maps.
 
The top panel shows comparison of the model with the ROSAT all-sky map \cite{snowden+1997}, where the interstellar absorption is calculated for a disc model of neutral gas. 
The middle panel shows a simulated three-color X-ray image in the 0.25, 0.75 and 2 keV compared with the eROSITA observation for the inner region of GSB \cite{predehl+2020}. 
The interstellar absorption has been corrected for using the HI survey data \cite{kalberla+2005}.
The simulations well reproduce the observed X-ray morphology of GSB and the increasing absorption and hardening of the spectrum toward the Galactic plane.   

Besides the bubbles and outflows, a giant HI hole has been known around the GC, which composes a bipolar conical cavity of neutral gas and is supposed to have been blown off by a galactic wind
\cite{lockman1984,lockman+2016}. 
The HI hole has its root in the Galactic plane, coinciding with the 3-kpc expanding ring \cite{sanders+1974,bania1980,cohen+1976}.
The cylindrical boundary of the HI hole makes a crater-like 
structure of radius 3 kpc around the GC, whose tangential directions coincides with the roots of the North Polar Spur and its counter spurs \cite{sofue2017}. 

The North Polar Spur (NPS) is the most prominent spur in the galactic halo in radio \cite{haslam+1982} and soft X rays \cite{snowden+1997,predehl+2020}. 
The radio NPS emerges from the galactic plane at G22+02 
\cite{sofue+1979}, where the spur exhibits maximum brightness and the sharpest cross section. 
This position exactly coincides with the tangential direction of the 3-kpc expanding ring in HI and CO lines \cite{sofue2017}.  

Energetic events in the GC associated with giant bubbles and shock waves, thus, exhibit various foot prints in the halo and disk from neutral ISM to high-energy emissions. 
In this paper we re-examine the morphological correlation between the 3-kpc crater and the GSB, looking for further evidence of the GC explosion in the HI and molecular discs using radio continuum, X-ray, HI and CO line surveys.  We, then, attempt to give a comprehensive explanation for various diverse features observed over the wide range of spectrum.

\section{X-Ray GSB absorbed by the disc}
 
\subsection{Extinction of X-ray spurs by the galactic disc}

\begin{figure} 
\begin{center}      
\includegraphics[width=8cm]{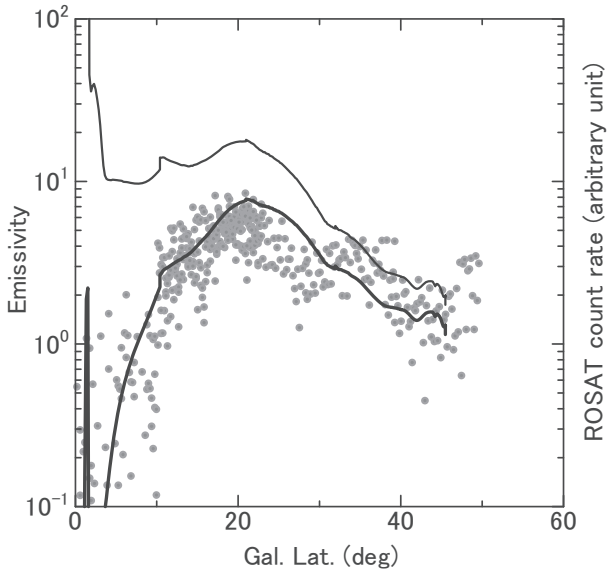}   
\end{center}
\caption{
ROSAT X-ray count rates at 0.75 keV along the NPS in arbitrary unit by dots (Snowden et al. 1997) compared with the intrinsic emissivity along the post shock from simulation of the starburst model by the thin line (Sofue et al. 2016), and with that after correction for the interstellar extinction due to the galactic disc by the thick line. Compare the relative latitudinal variations, as the absolute values are arbitrary.
}
 \label {fig2}    
\end{figure}

The top panels of Fig. \ref{fig1} demonstrate the heavy extinction of the X-rays from the GC and halo by the galactic disc.
Fig. \ref {fig2} shows the latitudinal variation of the X-ray intensity at 0.75 keV as observed by ROSAT \cite{snowden+1997}. 
The X-ray intensities along the spurs are significantly depressed at latitudes lower than $|b|\lesssim 10\deg$.
Such depression can be understood as due to the following two kinds of bounded features at low latitudes.
 
(a) Absorption-bounded cut due to interstellar extinction by heavy elements, whose optical depth is proportional to the line of sight depth of the disc obeying the sec $|b|$ law \cite{sofue1994,sofue2015,lallemant+2016} (Fig. \ref {fig2}).
This cut occurs at $|b|<\sim 10-20\deg$ by the local HI disc.

(b) Emission-bounded cut due to suppression of the intrinsic emissivity at low latitudes according to decreasing temperature  in the galactic disc, where neutral gas is dominant at temperatures below $<\sim 10^4$ K as predicted by simulations \cite{sofue+2016} (Fig. \ref{fig3}).
This cut occurs at altitudes as low as $|z|\sim <200$ pc, or $|b|<\sim 2\deg$, which is difficult to detect even it is present because of the strong absorption by the foreground disc gas as above. 

In either case, the observed apparent cut of soft X-ray brightness near the galactic plane makes contrast to the continuous radio continuum ridge of the NPS intersecting the disc. 

The latitudinal variation of the X rays due to interstellar extinction has been calculated using the hydrodynamical simulation as shown in Fig. \ref{fig1}. 
Fig. \ref {fig2} compares the expected variation of intrinsic emissivity along the simulated post shock (the densest inner shell of the simulation) by a thin line  with that after correction for the interstellar extinction by a thick line.
Here, the galactic disc is expressed by a plane parallel gas layer of thickness of 100 pc having absorption coefficient of $5\times 10^{-3}$ pc$^{-1}$, or $\tau=1$ by 200 pc in the galactic plane.
We plot the observed ROSAT intensities along the NPS by grey dots, which are well fitted by the simulation with the disc extinction.

From the strongly extinct nature near the Galactic plane as proved by Fig. \ref{fig1} and \ref {fig2}, we may argue that the X-ray maps are not appropriate for studying the physics of GSB interacting with the disc.  
  
\subsection{Temperature variation with height}

\begin{figure} 
\begin{center}      
\includegraphics[width=8cm]{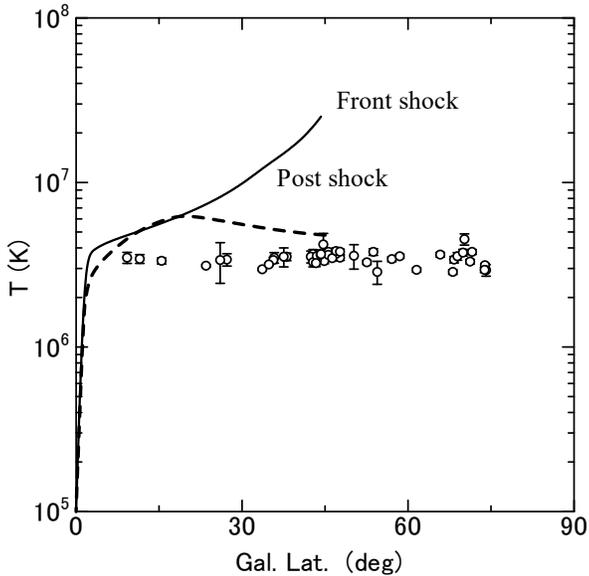}  
\end{center}
\caption{Gas temperature against latitude ($={\rm atan} (z/8 {\rm kpc})$) at the front shock (full line) and post shock (dashed line) from simulation of the starburst model,  
compared with observed X-ray temperatures from SUZAKU (circles) (Kataoka et al. 2013; Tahara et al. 2015; Akita et al. 2018) and HaloSat (LaRocca et al. 2020). 
}
 \label{fig3}   
\end{figure}

On the other hand, X-ray maps at high latitudes may be used to represent and study various physical properties of the GSB. 
Fig. \ref{fig3} shows gaseous temperature along the NPS plotted against the galactic latitude as observed with SUZAKU  \cite{kataoka+2013,tahara+2015,akita+2018} and HaloSat \cite{larocca+2020}.
In all the data plotted here,  the X-ray spectra are modeled by a same three-component plasma model: APEC1 + Wabs $\times$ (APEC2 + PL), where Wabs represents the Galactic absorption, APEC1 is an unabsorbed thermal component that represents the Local Bubble (LB) emission, APEC2 is an absorbed thermal component that represents the shocked halo gas like NPS, and PL is the contribution from Cosmic X-ray background. Note that the temperature of APEC2  is nearly constant at $kT\sim 0.3$ keV, or $T\sim 3\times 10^6$ K (but see a slight different version of model fitting in which APEC2 is further divided into $kT$ $\simeq 0.2$ and 0.4~keV components corresponding to unshockd and shocked halo gass \cite{miller+2016}). 
According to the bipolar-hypershell model, the high temperature can be reasonably attributed to heating of the halo gas by explosive event in the GC associated with a giant shock wave expanding at $\sim 200-300$ \kms {\cite{sofue1980,sofue+2016,sarkar2019,zhang2020}}. 

According to the simulation near the Galactic plane at $z<\sim 0.2$ kpc, the temperature is expected to be significantly lower than that in the halo due to the stronger deceleration of the shock wave inside the disc, causing decrease in the temperature, and so the cooling rate and emissivity of X rays (figure \ref{fig3}).  
However, the current X-ray observations cannot measure the intrinsic behavior at lower-latitudes than $|b|\le 10\deg$ because of the strong absorption by the foreground galactic disc as in figure \ref{fig2}.

As shown in Fig. \ref{fig3}, observed X-ray gas temperature of $kT$ $\simeq$ 0.3~keV is surprisingly uniform in the halo against latitude, and is roughly consistent with the "post shock" temperature predicted from the simulation of the starburst origin model. 

On the other hand, temperature of "front shock" is generally much higher.
In fact, MHD simulation predicts that the shock velocity 
gets even faster, sometimes more than {$\simeq$1000 km s$^{-1}$}, {or 
$kT$ $\simeq$ 1~keV}
at high $b$ owing to substantial decrease of the halo gas density 
where the shock propagates. 
Note that the cooling time scale 
of the X-ray emitting gas is expressed as 
 \begin{equation}
   t_{\rm cool} \simeq  50 \left (\frac{v_{sh}}{300~{\rm km~s^{-1}}}\right)^{\alpha}\left(\frac{n}{0.01~\rm{cm^{-3}}}\right)^{-1} ~\rm{Myr}, 
 \end{equation}  
for halo gas of subsolar metallicity $Z$ $\simeq$ 0.2 $Z_{\odot}$
 \cite{draine2011,kataoka+2021},
 where $\alpha=3.4$ at lower temperature than $kT\sim 1$ keV {or velocity $v_{\rm sh}\sim 1000$ \kms} 
 and $\alpha=1$ beyond.
 Assuming $v_{sh}$ $\sim$ 1000 km s$^{-1}$ 
 and $n$ $\simeq$ 0.001 cm$^{-3}$ for front shock at high $b$, we obtain
 $t_{\rm cool}$ $\simeq$ 1.5 Gy, that is much longer than dynamical scale for the bubbles' formation. 
 Hence the gas heated by front shock can be effectively adiabatic during the expansion.  
 
The emissivity of the X-ray emitting gas scales as $\propto$ $n^2$, thus most of the observed emission comes from post-shock region rather than front shock, where $kT$ $\simeq$ 0.3~keV plasma dominates. 
 The cooling time is then $\sim 12$ My, still long enough for the gas being regarded to be adiabatic. 

The observed offset between radio and X-ray NPS is attributed to the position of the front shock and post shock where the radio and X-ray emission becomes strongest. In fact, our MHD simulation \cite{sofue+2016} predicts $\sim$ 1~kpc offset at $z$= 4~kpc (or $b$ $\simeq$30$^{\circ}$), that corresponds to $\simeq$ 7~deg. This is roughly consistent with the observational radio/X-ray offset of $\simeq$ 5~deg at the brightest part of the NPS \cite{kataoka+2021}. 

\subsection{Relieved X-ray maps and innermost X-ray spurs}

\begin{figure} 
\begin{center}   
\includegraphics[width=7.5cm]{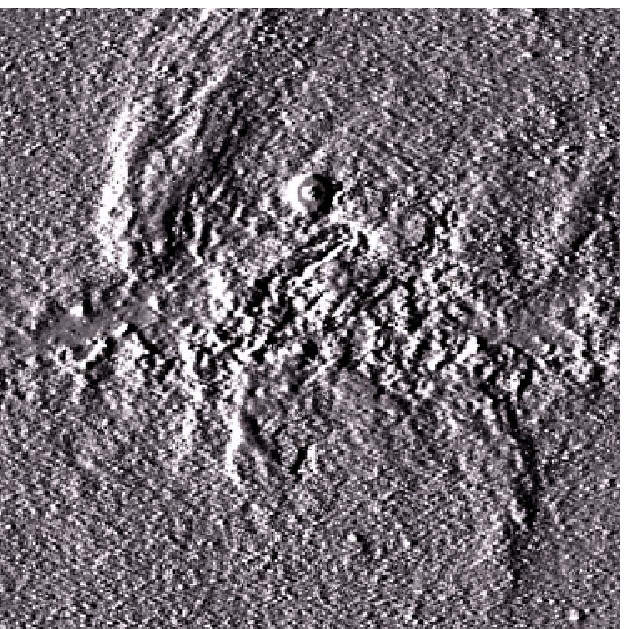} 
\end{center}
\caption{Relieved X-ray image produced from eROSITA all sky map at 1 to 2.3 keV using the color image from (Predehl et al. 2020).  
} 
\label{fig4}  
%\end{figure}   
%\begin{figure} 
\begin{center}    
\includegraphics[width=7.5cm]{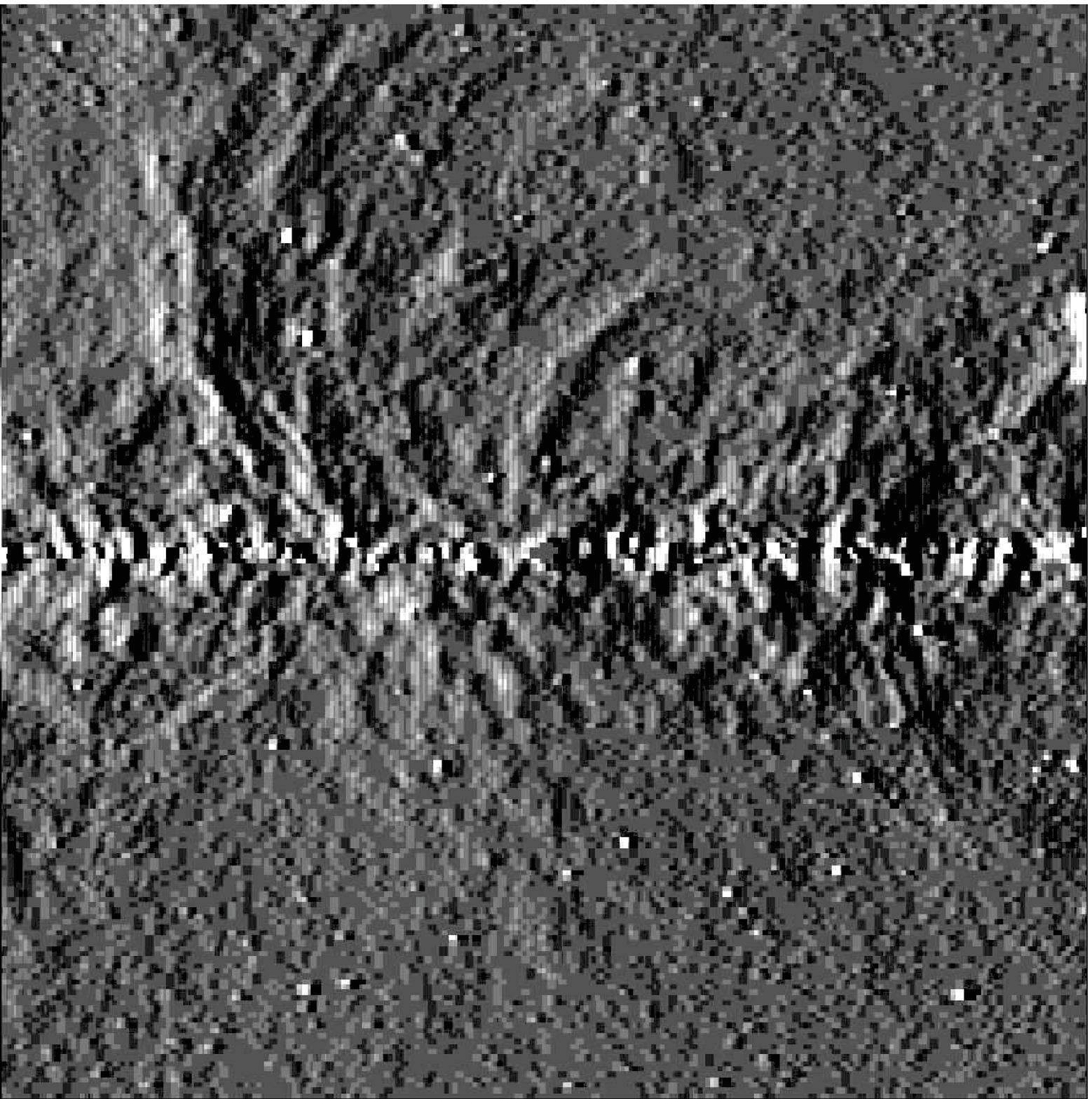}   
\end{center}
\caption{Horizontal-relieved map of $100\deg \times 100\deg$ region around GC at  1420 MHz (Reich et al. 2001).
}
\label{fig5} 
\end{figure}

In Fig. \ref{fig4} we show a relieved map of soft X-ray intensity at 1 to 2.3 keV as produced from the eROSITA all sky map \cite{predehl+2020}.
The relieving was applied by shifting the image in the longitude direction by a couple of angular resolution, then  subtracting it from the original image. 
By this simple procedure, the strong and relatively uniform galactic background is subtracted and small scale structures are enhanced. 
Note, however, that the images may not be used for quantitative measurement of the intensity, but is useful only for morphological study of the edged structures.

In the outer region at $|l|\sim > 20\deg$, X-ray spurs are coherently aligned along the radio spurs with significant displacements toward inside the bubbles \cite{kataoka+2021}.
Symmetric to the North Polar Spur (NPS) with respect to the GC, a bright X-ray southern spur at $\sim$G$-30-20$, drawing a clear arc, is associated with the South Polar Spur West (SPS-W) in radio continuum. 
An X-ray spur at $\sim$G$-30+20$ is associated with the radio spur NPS-W, and the X-ray spur at  $\sim$G$+30-30$ corresponds to SPS-E in radio. As a whole, both the radio and X-ray GSB compose an
\OmegaGSB
shape, making a symmetric dumbbell structure around the GC.
Note, however, that they do not draw an {\Large 8} shape crossing at the nucleus.
However, low-latitude extension below $|b|\sim <5\deg$ of the X-ray spurs of the GSB and their connection to the Galactic disc at $l\sim \pm 20-30\deg$ cannot be studied from the maps because of the strong absorption by the disc gas.

On the contrary, in a more central region at $|l|\sim < 20\deg$, we find numerous vertical X-ray structures, which include the near-GC spurs at G+10-05 (SE Claw) and G-05+05 (NW Clump) \cite{kataoka+2015} and the X-ray wind horns \cite{bland+2003}.
It is interesting to point out that these vertical features are located inside the HI wind cone. 
High-velocity ionized gas observed below the GC \cite{savage+2017} could be related to these near-GC X-ray spurs. 

\subsection{Origin of energy injection: AGN or starburst?}

The GC explosion model of the large scale bubbles seems to be  promising, but origin of energy injection is yet to be understood, 
either by an AGN-like outburst or from nuclear starburst  in the GC. 
In general, AGN-like models assumes that energy is released as a single 
point-like explosive event with a short time-scale in the nucelus, whereas starburst assumes that the energy is continuously released and injected to the GC over Myr timescale. While the total amount of energy required is almost the same, AGN model predicts younger bubble ages of 1$-$3 Myr owing to fast expansion velocity of 1000-10000 km s$^{-1}$ \cite{guo2012}. Such high velocity is apparently inconsistent with the observed X-ray temperature as discussed above, but situation can be significantly moderated by assuming constant AGN activity or continuous energy injection over Myr, leading to an expected plasma temperature of $kT$ $\simeq$ 0.4~keV \cite{zhang2020}. 

In this context, starburst models more naturally explain observed X-ray temperature and overall morphology of bubbles owing to  continuous star forming activity over $\gtrsim$10~Myr \cite{sarkar+2016,sofue+2016,sarkar2019}. 
In this scenario, the shock velocity and temperature in the highest compression region in the post shock wave is much smaller than that at the shock front by AGN models.
The heated plasma can be gradually dissipated and/or cooled to form quasi-isothermal bubbles as observed in X-rays. 
However, it has been argued that the observed 
subsolar metallicity in the bubbles, $Z$ $\simeq$ 0.2 $Z_{\odot}$, seems to be difficult to be understood within the framework of starburst models \cite{inoue+2015}. 

It seems that difference between AGN and starburst models simply depends on the energy injection time scale, rather than physical origin itself. 
Thus an alternative idea could explain the isothermal nature by assuming recurrent and continuous AGN activity as a source of energy injection over Myr time-scales. 
In fact, such recurrent activity are often observed in number of radio galaxies wherein multiple/mini radio lobes are found in quite different spatial size (e.g., Centaurus A: \cite{israel1998} for a review). 
The X-ray chimney, which looks like mini-bubbles, recently discovered in the GC is also collaborating this idea \cite{ponti+2019}. 
It is however difficult to clearly distinguish AGN and starburst scenario without detailed measurement of physical conditions of X-ray emitting plasma, such as the gradient of temperature, 
metallicity and ionization parameters against both 
latitude and longitude directions.

%%%%%%%%%%%%%%%%%%%%%%%%%%
 
\section{Radio GSB intersecting the disc}

\subsection{Radio bubbles}

Fig. \ref{fig5} shows a relieved image of radio continuum emission at 1420 MHz for a $100\deg \times 100\deg$ region centered on the GC made from the Bonn-Stockert-Argentine all-sky survey \cite{reich+2001}.
Almost identical features are observed in the relieved all-sky maps from the all-sky radio surveys at 408 MHz \cite{haslam+1982} and 2300 MHz \cite{jonas+1998}.

The north-eastern edge of the GSB, or the North Polar Spur (NPS), is visible as the brightest radio and X-ray ridge near $l=30\deg,\ b=+20\deg$ (hereafter, G30+20).
The western counterpart to NPS (NPS-W) is recognized near G-30+20 extending from the galactic plane to positive latitude. 
Their southern counterparts are recognized around G30-20 and G-30-20, which were called the South Polar Spurs East and West (SPS-E, SPS-W). These four spurs (NPS, NPS-W, SPS-E and SPS-W) compose a bipolar double-horn structure symmetric around the galactic plane and the rotation axis of the Galaxy. 

\subsection{Intersection with the galactic plane}

The radio continuum spurs composing the GSB extend toward the galactic plane without significant absorption, and merge with the complex disc composed of numerous radio sources and extended structures.
The northern radio spurs apparently stop at the Galactic plane, and appear to be connected to the southern spurs, although the details at $|b|\sim <1-2\deg$ are not well established in the figures mainly because of the contamination by the fore- and background radio sources through the Galactic disc. 

Among the spurs composing the GSB, the NPS is most prominent about the relation to the Galactic disc.
The extension of the radio NPS toward the Galactic plane has been studied using background-filtered continuum maps observed with the Effelsberg 100-m telescope 
\cite{sofue+1979,sofue2000}. 

\begin{figure}
\begin{center}     
\includegraphics[width=8cm]{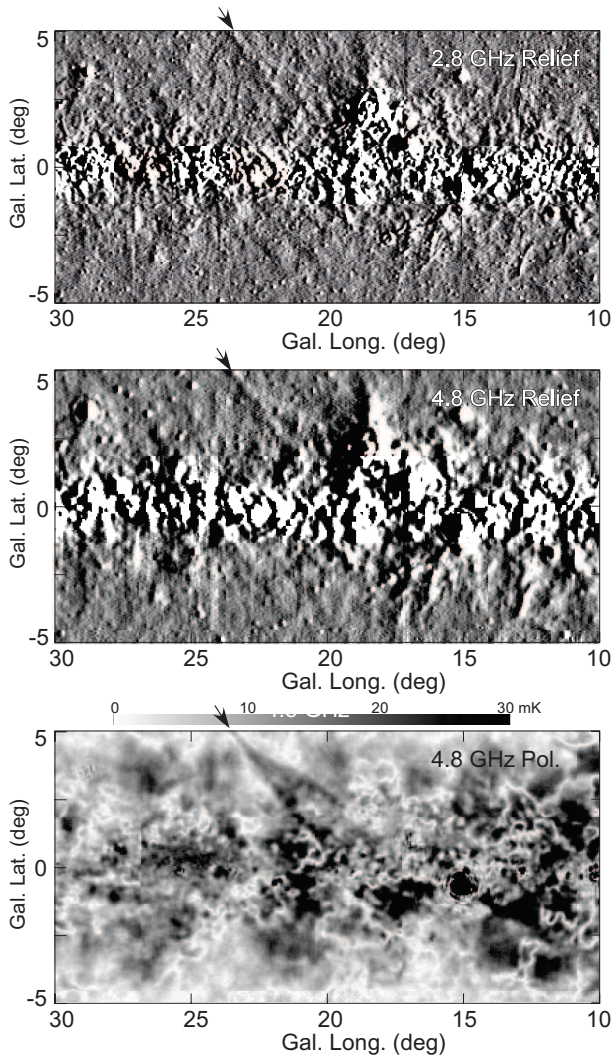}  
\end{center}
\caption{[Top] 2.8 GHz relief map from Bonn-100 m telescope survey (Reich et al. 1990) near the root of the NPS.
The arrows points the extension of the NPS.
[Middle] 4.8 GHz relief from Urmqi 25-m telescope survey  (Sun et al. 2010). 
[Bottom] Same, but polarized intensity. Arrows point the NPS ridge. A bright source at $\sim$G19+03 is a foreground HII region, which is not visible in polarization. 
}
\label{fig6} 
\end{figure}

 In order to see the intersection of NPS with the galactic plane in more detail, we show in Fig. \ref{fig6} a relieved 2.8 GHz map made from the survey with the 100-m telescope at a resolution of $4'.3$ \cite{reich+1990} and relieved 4.8 GHz map from the Urmqi 25-m telescope at a resolution of $9'.5$ \cite{sun+2011}.   
In the bottom panel, we show linearly polarized intensity map at 4.85 GHz \cite{sun+2011}, indicating that the spur's non-thermal nature of synchrotron origin. 
The polarization angle of the E vector along the spur lie at PA$\sim 130\deg$, indicating that the B (magnetic) vector is at PA$\sim 40\deg$, or the magnetic field is aligned along the spur, if the Faraday rotation is negligible.  
 
At very low latitudes of $|b|\le \sim 1\deg$, the radio continuum NPS, both in intensity and polarization, shows no signature of extension across the galactic plane. 
The fact that the NPS terminates at G21+01 at its brightest and sharpest end indicates that the NPS has its origin in the Galactic disc.

\subsection{Magnetic fields}

Besides the transparency for studying the morphological detail even through the galactic disc, an advantage to use radio continuum emission is its direct relation to the magnetic field in the shell. 

 If cosmic rays (CR) are produced by turbulent acceleration in the compression site of magnetic fields, we may assume equipartition between CR  and magnetic energy densities.
Then, the magnetic strength can be estimated using the relation \cite{sofue+2019mag},
\begin{equation}
B\sim 3.1\times 10^8 (\nu/1\ {\rm GHz})^{-1/7}\epsilon^{2/7} 
\ \mu{\rm G},
\end{equation} 
where $\nu$ [in GHz] is the radio frequency and $\epsilon =4 \pi \int  \Sigma_\nu  d\nu/L\sim 4 \pi\nu \Sigma_\nu/L$ [in erg cm$^{-3}$ s$^{-1}$] is the radio emissivity  for line-of-sight depth $L$ of the emission region. 
Here, $\Sigma_\nu=2 k {\Tb}_\nu/\lambda^2$ is radio surface brightness at $\nu$, ${\Tb}_\nu$ the brightness temperature, and $\lambda$ the wavelength.   

%Out[494]=\!\(2.30000000000000071`*^9\)
%Out[508]=0.5
%Out[525]=\!\({1.51937166519459321`*^-29}\)
%Out[495]={1.59612}
%Out[496]=\!\({1.01365508818958094`*^-13}\)
%Out[497]=\!\({2.115531882428483`*^8}\)

%Out[498]=\!\(1.40000000000000035`*^9\)
%Out[513]=1.4
%Out[531]=\!\({9.59450676257343459`*^-30}\)
%Out[499]={1.50254}
%Out[500]=\!\({8.98275107675291373`*^-14}\)
%Out[501]=\!\({2.96879428129531532`*^8}\)

%Out[502]=\!\(4.08000000000000095`*^8\)
%Out[518]=40
%Out[537]=\!\({6.78501779758235912`*^-30}\)
%Out[503]={1.62304}
%Out[504]=\!\({1.04813296151753432`*^-13}\)
%Out[505]=\!\({4.89845114229337141`*^8}\)

%Out[506]=\!\({4.79999999999999893`*^-12}\)
 
Inserting observed excess in brightness temperature of 
${\Tb}\sim 0.5$ K at 2.3 GHz \cite{jonas+1998},
$\sim 1.4$ K at 1.4 GHz \cite{reich+2001},
and $\sim 40$ K at 408 MHz  \cite{haslam+1982} 
toward the NPS ridge at $b\sim 30\deg$, we obtain volume emissivity of
$\epsilon \sim 1.5\times 10^{-29}$, 
$\sim 0.96\times 10^{-29} $, and
$\sim 0.68\times 10^{-29}$ \ergccs, respectively.
These values yield magnetic strength of 
$B\sim 1.6$ \muG 
for an assumed line-of-sight depth of $L \sim 0.5$ kpc. 
Then, the magnetic energy density, which is equal to CR energy density, is 
$u_{\rm mag}=u_{\rm CR}\sim B^2/8 \pi  \sim 1.0\times 10^{-13}$ ergs cm$^{-3}$.
This is smaller than that of the X-ray emitting thermal gas with $kT \sim 0.3$ keV and $n\sim 10^{-2}$ cm$^{-3}$, having $u_{\rm thermal} \sim 5\times 10^{-12}$ ergs cm$^{-3}$.
The decay time of CR electrons emitting at 1.4 GHz is estimated to be $t_{\rm CR \ decay}\sim u_{\rm CR}/\epsilon \sim 0.4$ Gy. 
This is longer than the dynamical time of the shock wave, $t_{\rm dyn} \sim 10$ My, but shorter than the cooling time of the thermal gas, $t_{\rm cool}\sim 1.5$ Gy. 
 
{It is noted that much more energy may be reserved in CR protons rather than electrons, but the emission from CR protons are hard to be observed. 
In fact, recent MHD simulation on the SNR shock suggests that energy density in 
the CR protons is several 
orders of magnitude larger than 
that of CR electrons, depending on the 
SNR age in Sedov phase \cite{uros+2018}. Moreover, given that only a small fraction of cold protons are picked up into the CR acceleration process, we should also 
consider hidden contribution of $cold$ 
protons as a reservoir of CR protons. 
Thus the substantial mismatch between CR electrons, magnetic fields and X-ray emitting gas may suggest that most of energy 
may be stored by protons 
in the front shock of the NPS and GSBs. 
}

Morphological comparison between X-ray and radio spurs on the sky has posed an interesting question about the displacement between the X-ray and radio edges, such that the X-ray shell is located significantly inside the radio shell \cite{kataoka+2021}. 
The large displacement may imply either that the frozen-in nature does not hold between the gas and magnetic fields in the GSB.
Alternatively, the acceleration of CR electrons is local and their life time is comparable with the GSB's evolutionary time scale, which is, however, marginal in view of the life time estimation in section 4.3.
 
%%%%%%%%%%%%%%%%%%%%%%%%%
\section{3-kpc crater as a footprint by GSB on the disc}  
 
\subsection{GSB's root enveloped by HI hole and 3-kpc crater}

\begin{figure} 
\begin{center}  
\includegraphics[width=8cm]{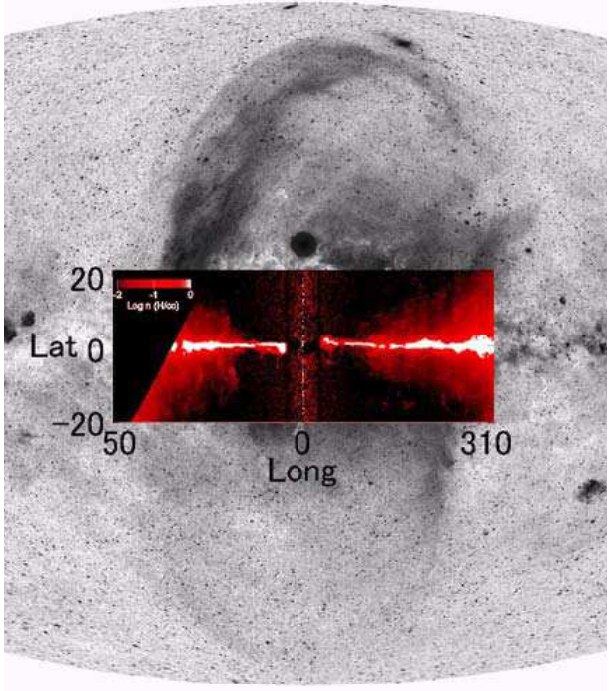}   
\end{center} 
\caption{ X-ray image by eROSITA (Predehl et al. 2020) at 1 to 2.3 keV compared with HI volume density ($n_{\rm HI} \propto \Tb$) at tangent velocities, showing the cylindrical cavity coincident with the bubbles (Sofue 2017).
}
 \label{fig7}  
\end{figure}  

We first overview the spatial relation of the GSB to the HI hole and 3-kpc crater in the Galactic disc and halo \cite{lockman1984,lockman+2016,sofue2017}.
In Fig. \ref{fig7} we overlaid a map of the HI volume density ($n_{\rm HI} \propto \Tb$) along the terminal velocity circle made from GASS HI survey \cite{mcclure+2009} on the eROSITA X-ray map \cite{predehl+2020}.
The low-latitude X-ray spurs composing the GSB approximately coincide in position with the inner wall of the HI hole (conical cavity). 
The inner HI disk is as thin as $\sim 1\deg$ inside 3 kpc, but the disc suddenly becomes thicker and denser at $l\sim 22\deg$, attaining a thickness of $\sim 10\deg$, exhibiting a crater structure.
%Figure \ref{fig7} 
 The HI wall composes vertical spurs extending to $b\sim \pm 10\deg$ (1 kpc), coinciding also with the radio spurs.
These facts agree with the idea that the NPS and its counter spurs compose a dumbbell-shaped shock front, which swept away the interior HI disk inside $\sim 3$ kpc.  

\subsection{Longitude-velocity diagrams}

\begin{figure} 
\begin{center}     
\includegraphics[width=6.5cm]{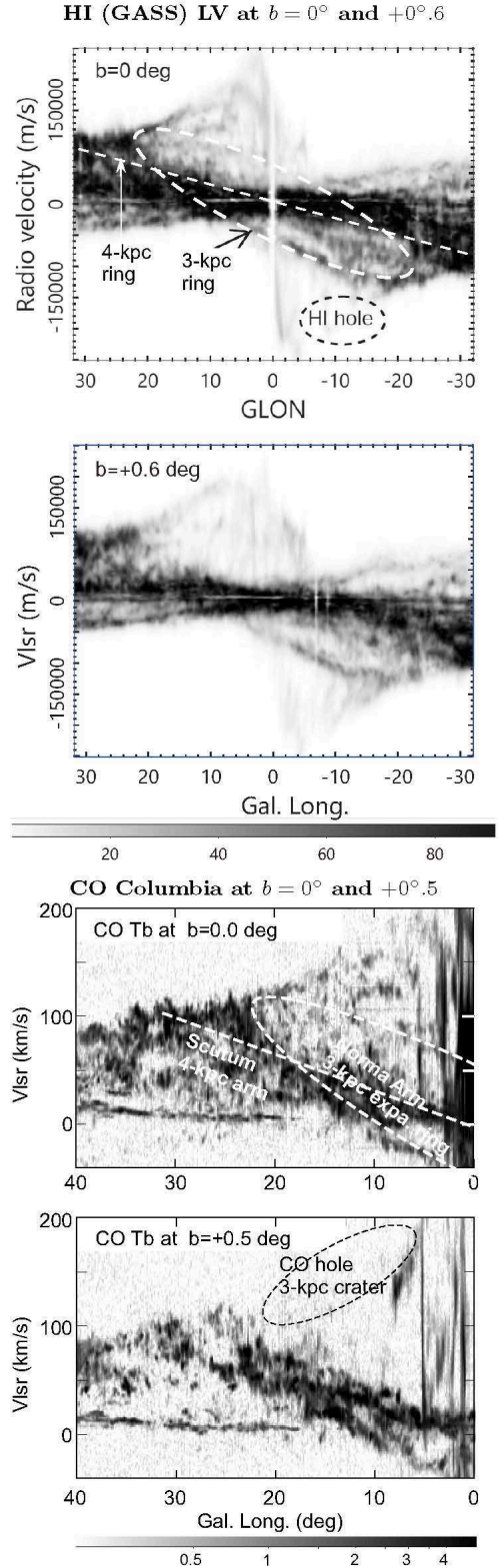}  
\end{center}
\caption{ 
 LV diagrams in the HI line at $b=0\deg$ and $+0.6\deg$ from GASS survey (McClure-Griffiths et al. 2009),  and CO line at $b=0\deg$ and $+0\deg.5$ from Colombia survey (Dame et al. 2001) . 
  The 3-kpc expanding ring is indicated by the dashed ellipse.}   
 \label{fig8}  
\end{figure}

Fig. \ref{fig8} shows HI line longitude-velocity (LV) diagrams made from the Parkes All Sky Survey (GASS) of the HI line emission \cite{mcclure+2009} at $b=0\deg$ and $0\deg.6$.
The 3-kpc expanding ring (HI crater's wall) is evident as the tilted ellipse of HI ridge in the LV diagram, which crosses $l=0\deg$ at positive and negative high velocities of $\pm 50-60$ \kms. 

\begin{figure} 
\begin{center}  
\includegraphics[width=8cm]{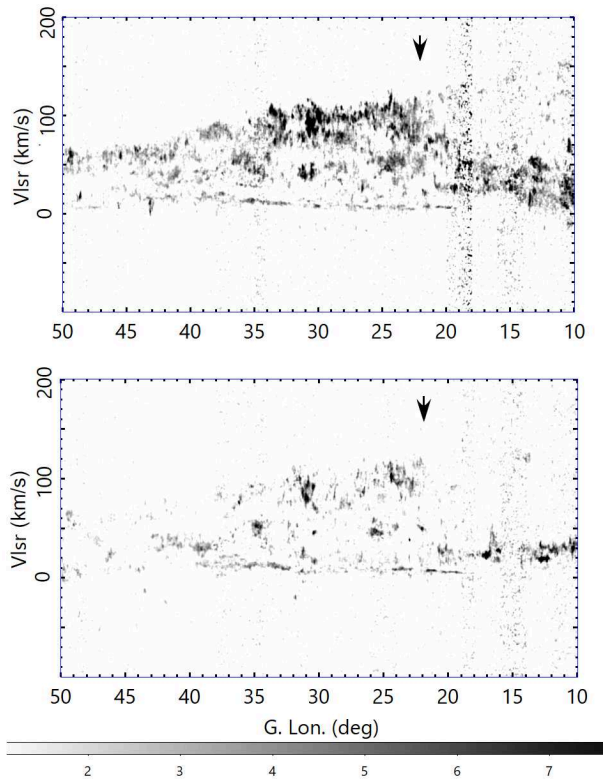}   
\end{center}
\caption{CO-line LV diagrams at $b=+0\deg$ (top) and $0\deg.5$ (bottom) from the FUGIN survey (Umemoto et al. 2017). Note the clear cut of terminal-velocity clouds at $l=22\deg$ as indicated by the arrows.} 
 \label{fig9}  
\end{figure}  

The negative velocity (approaching) arm is recognized at $b\sim \pm 0\deg.5-0\deg.6$ both in CO and HI LVDs, while positive velocity arm is fainter at $\pm 0\deg.5$.
In the off-plane regions at $b=\pm 0\deg.5-0\deg.6$ at $|l|\sim \le \sim 20\deg$, there appears a vacant region of tangent-velocity gases, marked as HI hole.
Such an LV behavior implies that the gases are empty inside 3 kpc at these latitudes.

Fig. \ref{fig9} shows LV diagrams of the \co emission at higher resolution from the FUGIN\footnote{Four-receiver system Unbiased Galactic Imaging survey with the Nobeyama 45-m telescope} \cite{umemoto+2017} CO survey around the northern tangent point of the 3-kpc ring.
The figure indicates that the upper layer of the molecular disc is sharply truncated at $b\sim 0\deg.5$ (65 pc).

\subsection{ Explosion vs bar origin} 

 The non-circular motion exhibited by the elliptical feature in the LV diagram in the 3-kpc arm has been interpreted in two ways.  

One idea is that the ring is expanding, accelerated by an explosion at the GC  
\cite{sanders+1974,bania1980,sofue1984}. 
The gas is approaching to the Sun in the near side at negative radial velocity of $-53$ \kms, while receding in the far side at positive velocity at $+56$ \kms \cite{dame+2008}. 
This motion produces a simple LV ellipse, as shown by
the dashed line in figure \ref{fig8}, which reasonably traces the
observed HI LV ellipse.
The total mass of the CO and HI rings is on the order
of $\sim 10^8 \Msun$, and their kinetic energy of the expanding motion at 50 \kms is $\sim 5 \times 10^{54}$ erg \cite{sofue2017}.
This requires only a small portion, $\sim 0.1$, of the total energy to drive the ring's expansion. 
The angular momentum problem may be eased, if the ring is a focusing wave front and not an accumulated ring \cite{sofue1977}.
 
 Another idea is that the non-circular LV feature is produced by a bar potential \cite{wein1992} according to the various families of "x" orbits \cite{conto1956}.
Simulations have shown complicated LV features, known as the parallelogram, whose orientation depends on the position angle, mass, axial lengths and ratio of the bar in a complex way
\cite{binney+1991,athana+1999,sormani+2015}.

We may also combine the two ideas in such a way that the halo structures are produced by the explosion, while the disc's LV feature by the bar.
The bar promotes accretion of the disc gas to GC, and triggers starburst and/or AGN, producing the GSB.
However, in this case, we must assume 1) that the exact coincidence of the 3 kpc arm with the roots of GSB and HI cylinder happened by chance, 2) that the gigantic explosion did not give any imprint on the disc, and 3) that the bar and flow orientation must be special to produce the observed simple LV oval.

Considering these, particularly the fact that the observed 3-kpc LV oval is fitted by a simple ellipse fairly well as in figure \ref{fig8}, not particularly requiring a parallelogram, we here attempt to explain the GSB, HI cylinder, 3-kpc ring, and the HI+CO crater structure by a single hypothesis of the GC explosion.
Note that the explosion model explains the 3D structures in the halo and disc, whereas the current bar simulations cannot touch upon the vertical structures.
A more sophisticated 3D model incorporating a bar and GC blast would be desirable, although it is beyond the scope of this paper. 

\subsection{Density and disc thickness jumps at 3 kpc crater}
 
\begin{figure} 
\begin{center}  
\includegraphics[width=8cm]{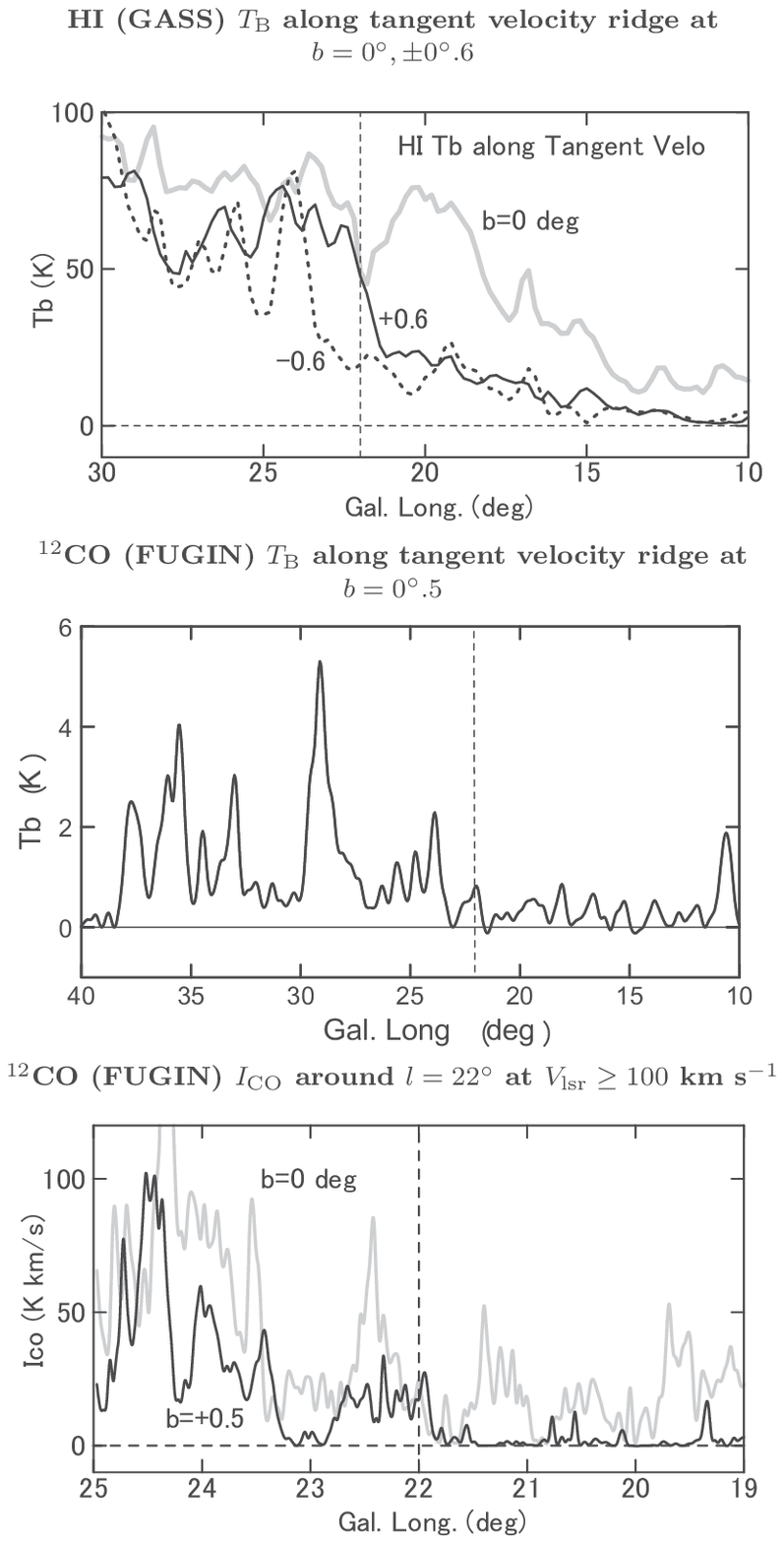}     
\end{center}  
\caption{[Top] Longitudinal variation of $\Tb$ of HI gas (GASS) at tangent points at different latitudes, $b=0\deg$ and $\pm 0\deg.6 (\pm 78 $ pc) . 
[Middle] FUGIN \co $\Tb$ at tangent points at $b=+0\deg.5$ (+65 pc).
[Bottom] FUGIN \co intensity at $\vlsr \ge 100$ \kms at  $b=0\deg$ and $0\deg.5\ z=+65$ pc.  
Note the jump of the gas density at $l\sim 22\deg$ corresponding to the 3-kpc crater, within which the HI and CO gas discs are truncated.
} 
\label{fig10}  
\end{figure}

Fig. \ref{fig10} shows longitudinal variations of the HI and CO line brightness temperatures $\Tb$ at the tangent-point velocities plotted against longitude in the galactic plane and at $b= \pm 0\deg.5-0\deg.6$.
We recall that $\Tb$ is proportional to the mean volume density in the line of sight depth through the tangent point gas as
\begin{equation}
n_i=\frac{dN_i}{dx}=\frac{dN_i}{dv} \frac{dv}{dx}=X_i \Tb \frac{dv}{dx},
\end{equation}
where $i$=HI or CO with $X_i$ being the conversion factors, $x$ and $v$ the distance along the line of sight and radial velocity, respectively.
The velocity gradient is on the order of $dv/dx\sim 5$ \kms kpc$^{-1}$ for a velocity gradient of $\sim 5$ \kms and a line of sight depth across the tangent-velocity gas of $\sim 1$ kpc.
For HI with $X_{\rm HI}=1.8\times 10^{18}$ H cm$^{-3}$ [K \kms]$^{-1}$ and $\Tb\sim 80$ K, we obtain a mean density of $n_{\rm HI} \sim 0.3$ H cm$^{-3}$.
For CO with $X_{\rm CO}\sim 2\times 10^{20}$ H$_2$ cm$^{-2}$ [K \kms]$^{-1}$ and $\Tb\sim 5$ K, we obtain $n_{\rm H_2}\sim 2$ H$_2$ cm$^{-3}$.

The $\Tb$ plots in Fig. \ref{fig10} show a step-like increase of the volume densities of HI and CO gas from lower to higher longitudes at $l=22\deg$.
In the Galactic plane ($b=0\deg$), the density jump of the molecular gas takes place from a finite but small value inside $22\deg$ to higher values.
On the other hand, the off-plane molecular gas is almost empty inside $=22\deg$, but suddenly attains finite values beyond this longitude.
Such a density profile indicates that the molecular gas disc is thin inside 22$\deg$, and is surrounded by a thicker disc beyond the longitude, making a wall-like distribution with a radius of 3 kpc.

\begin{figure} 
\begin{center}     
 \includegraphics[width=8.5cm]{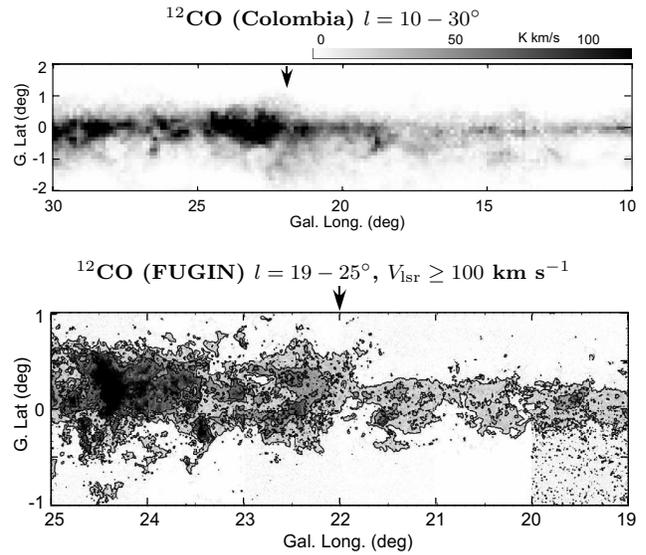}  
\end{center} 
\caption{
[Top] \co $\Ico$ maps integrated at $\vlsr \ge 60$ \kms from Columbia survey (top), and  
[Bottom] $\Ico$ from FUGIN survey integrated at $\vlsr \ge 100$ \kms. 
Contours are at 10, 30 and 50 K \kms.
Note the step-like increase of the disc thickness at $l\sim 22\deg$ as traced by the lowest contours, indicating truncation of the upper layer inside 3 kpc.
} 
\label{fig11} 
\end{figure}

Cross section of the gas distribution in the 3-kpc crater can be more directly displayed by channel maps in the HI and CO-line emissions at the terminal velocity.
Fig. \ref{fig11} shows $\Ico$ maps integrated at $\vlsr \ge 60$ \kms from the Columbia CO survey, and the same but at $\vlsr\ge 100$ \kms from the Nobeyama FUGIN survey around $l\sim 22\deg$.
A sudden jump of the thickness of CO disc is recognized at $l=22\deg$ such that the thickness by FWHM (full width at half maximum) of $\sim 0\deg.5$ at $l \le 22\deg$ to $\sim 1\deg.5$ at $l > 22\deg$.
 
The sudden change of the thickness revealed in the FUGIN CO map suggests that the upper layer of the molecular disc has been sharply truncated by sweeping mechanism inside $l\le 22\deg$ due to a shock wave from the GC or a horizontal wind blowing in the halo. 

\begin{figure} 
\begin{center}  
\includegraphics[width=8.5cm]{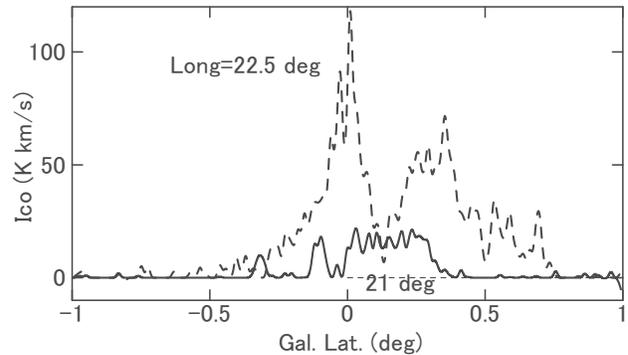}   
\end{center}
\caption{Latitude cross sections of $\Ico$ ($\vlsr\ge 60$ \kms) at G21 and G22.5 around the 3-kpc ring from FUGIN, showing drastic change of the vertical extent of the molecular disc from inside to outside the 3 kpc ring at $l=22\deg$.  }
 \label{fig12}  
\end{figure}

The thickness variation may be more clearly seen by comparing vertical cross sections at different longitudes outside and inside of the 3-kpc arm at G22. 
Fig. \ref{fig12} shows the cross section at G22.5 compared with that at G21. The former shows a wide extent of the disk, having an outskirt to high latitudes. On the other hand, the cross section at G21 shows a plateau-like distribution, sharply cut at $\pm 0\deg.25$ from the centroid, suggesting truncation of the outskirt. 

Thus, the horizontal (longitudinal) and vertical (latitudinal) variations show that the density and thickness of the galactic gas disk, both in HI and CO, increase suddenly at $l\sim 22\deg$. 
The variation is more step-like in the off-plane region at $|b|\sim 0\deg.5$ than in the galactic plane.  

From Fig. \ref{fig12} the molecular disk thickness inside G22 is as thin as $\pm 0\deg.25$ (30 pc) and increases to $\sim 1\deg$ outside $22\deg$. 
This indicates that the galactic disc of neutral gas (HI and CO) has a crater structure, having a hole inside G22 above the disk at $|z|>\sim 30$ pc ($0\deg.25$).
As shown in Fig. \ref{fig10}, the HI hole's wall has a radius of $r=3$ kpc and thickness $\sim 260$ pc (rising full width of $\sim 2\deg$ measured at $b=0\deg.6$).
On the other hand, the CO hole has a much sharper wall with thickness of $\sim 30$ pc ($\sim 0\deg.2$ at $b=0\deg.5$). 

\subsection{Formation of the crater and energetics}

\begin{figure} 
\begin{center}  
\includegraphics[width=7cm]{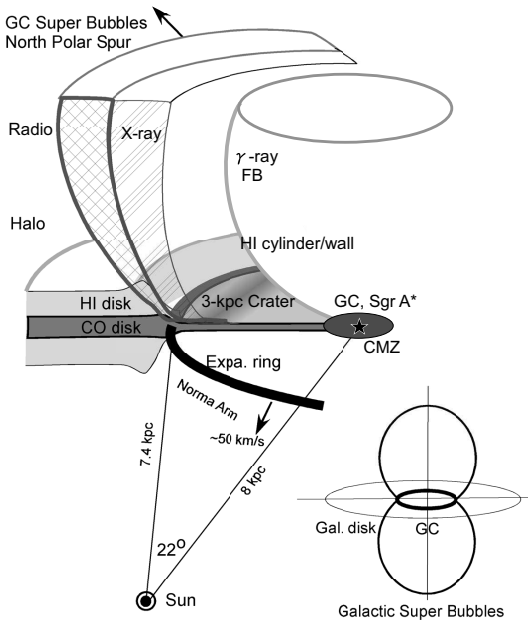}  
\end{center} 
\caption{Schematic illustration of the relation of GSB (NPS) to the GC and 3-kpc expanding ring (Norma Arm) and crater.}
 \label{fig13}  
\end{figure}

It has been shown that the root of the GSB can be traced by radio continuum ridges intersecting the Galactic plane.
The NPS root was shown to exactly coincide with the 3-kpc expanding ring composed of HI and molecular gases. 
Fig. \ref{fig13} illustrates a schematic view of the interaction of the GSB with the HI and CO disc based on the observed data shown in the previous subsections. 
In the upper halo, the GSB makes giant shells with the radio continuum being located outside X-rays \cite{kataoka+2021}, and is enveloping the $\gamma$-ray bubble \cite{su+2010}. 
The upper layer of the galactic disc is swept away and accumulated in the wall of the 3-kpc crater, creating the HI and CO rings.
By the focusing effect of the shock wave on the disc \cite{sofue1977,sofue2020b}, the front near the disc makes a compressed ring with the expansion velocity at $\sim 50$ \kms.
However, the thin and densest area near the Galactic plane at $|b|\lesssim 1\deg$ remains almost unperturbed because of the rapid deceleration of the shock wave in the plane inside 1-2 kpc from the GC.

Using the observed brightness temperatures and three-dimensional distribution expected from the apparent extent on the sky and assumed distance, the mass of the 3-kpc ring has been estimated to be $1.5\times 10^8 \Msun$ \cite{sofue2017}.
The corresponding kinetic energy is estimated to be on the order of $\sim 5\times 10^{54}$ ergs for the expanding motion at 50 \kms \cite{sofue2017}.
This energy is an order of magnitude smaller than the currently estimated total energy required for the formation of the GSB of $\sim 10^{55-56}$ ergs.
Hence, energetically, it is reasonable to assume a common origin for the formation of the GSB and the 3-kpc expanding ring with the HI and CO crater in the galactic disc.  

%%%%%%%%%%%%%%%%%%%%%%%%%%%%%%%%%%%%%
\section{Discussion}

\subsection{Unified view of the GSB in X, radio, HI and CO}

We have re-visited the GSB using the all-sky X-ray and radio continuum surveys as well as HI and the newest CO-line data of the galactic disc.
From the wide coverage of the observed energy ranges as well as by the spectral line information, we are now able to obtain a unified view of the GSBs as follows.

The X-ray data are crucial to derive thermal and kinematic properties of the GSB as a shock wave from the GC propagating through the halo gas.
Gaseous temperature can be used to calculate the shock wave velocity $v$, which is related to the injection energy $E_0$ at the GC and accumulated total mass $M$ of the shell by $E_0\sim 1/2 M v^2$ for an adiabatic shock wave as reasonably assumed from the sufficiently long cooling time. 
From X-ray spectroscopy, we know that $kT \sim 0.3$ keV (Fig. \ref{fig3}), and the intensity yields thermal electron density on the order of $n\sim 10^{-2}$ cm$^{-3}$ for assumed radius $\sim 4 $ kpc and thickness $\sim 1$ kpc.
These estimations lead to $M\sim 2\times 10^8 \Msun$ and $E_0\sim 4\times 10^{55}$ ergs.
Under the adiabatic assumption, the age of the GSB is approximately estimated from the current expansion velocity as $t\sim 0.6 r/v\sim 11$ My.

Although X-ray data are useful for discussing the shock wave propagation and energetics, they strongly suffer from the interstellar extinction by heavy elements in the galactic disc, where the optical depth steeply increases toward the galactic plane as $\propto {\rm sec}\ |b|$.
This results in the dark absorption belt along the galactic plane in the all-sky X-ray maps, making it difficult to obtain information about the interaction of the high-temperature gas with the galactic disc.

Such difficulty can be eased by observing radio continuum emission that can be safely assumed to be transparent in the whole Galaxy including the disc. We have thus showed that the roots of the GSB are connected to the tangential directions of the 3-kpc expanding ring as revealed in the HI and CO line emissions.

The radio ridge of the NPS was shown to be extending straightly toward the galactic plane, getting sharper and brighter and terminating at $b\sim 1\deg$ after merging into the bright disc emission.
From the intersection of the GSBs' spurs at longitudes of $\sim \pm 20\deg$, we showed that the whole GSB structure composes an \OmegaGSB shape.

The distribution of HI and molecular gases along the tangent-velocity circle showed that the ring composes a crater-like structure in the disc around the GC.
The 3-kpc ring and crater wall are expanding at $\sim 50$ \kms as inferred from the LV diagrams.
Kinetic energy of the expanding motion of the ring is estimated to be on the order of $5\times 10^{54}$ ergs.
Namely, a small portion of the total energy required to drive the whole GSB ($10^{55-56}$ ergs) is sufficient to drive the expanding ring and to create the crater.
 
We emphasize that the GC explosion model explains the coherent 3D structures in the disc and halo around 3 kpc, which include the sudden increase in the disc thickness at $l\sim 22\deg$ beyond horizontal truncation of the upper layer inside 3 kpc and the cylindrical HI wall around the central cavity as well as the non-circular kinematics revealed in the LV digrams.   

\subsection{Galactic-scale feedback}

Finally, we will comment on the implications of the interaction between the GSB and the galactic disc, assuming that the expanding ring of the gas disk is due to a GC explosion.
Current research of GSB in the literature has been devoted to discussing its impact on the galactic halo.
However, in the following we highlight the impact on the disc and its activity.
This may somehow akin
to discussing the impact of an explosion on the earth not only on the atmosphere, but also on structures on the ground.

The shock wave velocity is variable with the gaseous density in such a way that the velocity decreases with the density.
Accordingly, the ray path of a wave is refracted due to varying propagation velocity in the disc and halo.
This causes a converging flow of the shock front toward the galactic disc, making a ring of focusing waves \cite{sofue2020b}.

If the wave amplitude is small enough, or it is a linear wave like sound and MHD waves, the focal length $f$ is approximately related to the scale height of the disc $h$ as $f\sim 2.2 \ h$ \cite{sofue1977,sofue2020b}.  
Thus, besides the large-scale energetic flow as huge bubbles into the halo, some fraction of the released energy at the GC propagates the galactic disc and lower halo, repeating focusing onto rings of radii $r_i\sim 2.2 \ i \ h $ with $i=1, 2, ...$. 

The 3-kpc ring could be one of such focal rings driven by the GC activity, although its amplitude is non-linear and the simple focusing law may not apply and the focal length would be longer than $2.2 \ h$.
Also, after passing a focal ring, the wave diverges again into the disc and halo, propagates further outward, repeating focusing.

It is, therefore, expected that some fraction of the released energy at the GC by starburst or AGN activity  is efficiently confined in the disc and lower halo,  and repeat convergence onto disc-wide focal rings.
We emphasize that, besides the GSB's effects on the halo and intergalactic space, the galactic-scale feedback of the GC activities back onto the disc is also important for considering the structure and evolution of the Galaxy \cite{sofue2020b}.
For example, convergence of the released energy from the GC, even though a fraction of the total amounting to $\sim 10^{55-56}$ ergs per $\sim 10$ My, would cause various dynamical effects in the disc, which include triggering of star formation by compression in focal rings, and generation of the interstellar turbulence.

\section{Summary}

We revisited the GC super bubbles (GSBs), being inspired by the eROSITA all-sky X-ray observations \cite{predehl+2020}, and discussed their origin as due to an explosion in the Galactic Centre.
With a particular emphasis of the interaction of the bubbles with the Galactic disc, we investigated the radio continuum, X rays, HI surveys and the newest high-resolution CO-line survey of the Galactic disc.
We showed that the roots of the bubbles coincide with the 3-kpc expanding ring and crater in the HI and CO discs of the Galaxy. 
The thus revealed 3D structure of the crater is explained by sweeping of the disc gas by the GSB as a footprint of the explosive event in the Galactic Centre. 
The origin of the GSB can be well modeled by a 
\OmegaGSB shaped shock wave from the GC with an explosive energy input of several $10^{55}$ ergs $\sim 10$ My ago.

%%%%%%%%%%%%%%%%%%%%%%%%% 
  
\section*{Acknowledgements} 
%The data analysis and computations were carried out on the data analysis computer system at the Astronomy Data Center of the National Astron. Obs. of Japan.
The authors are indebted to Prof. A. Habe for the hydrodynamical simulation for Fig. 1 as reproduced from our paper in 2016.
 J.K. acknowledges the support from  JSPS  KAKENHI  Grant  Number JP20K20923.
 
\section*{Data availabivity} 
The CO line data were taken from the FUGIN CO-line survey at url: https://nro-fugin.github.io/. 
The eROSITA X-ray image was taken from the url: https://www.mpe.mpg.de/7461761/news20200619.

%%%%%%%%%%%%%%%%%%%%%%%%%

\end{document}